# BeWith: A Between-Within Method to Discover Relationships between Cancer Modules via Integrated Analysis of Mutual Exclusivity, Co-occurrence and Functional Interactions

Phuong Dao[1,*], Yoo-Ah Kim[1,*], Damian Wojtowicz[1], Sanna Madan[2]
Roded Sharan[3#], and Teresa M. Przytycka[1#]

[1] National Center of Biotechnology Information, National Library of Medicine, NIH, Bethesda, MD, USA
[2] Department of Computer Science, University of Maryland, College Park, MD, USA
[3] Blavatnik School of Computer Science, Tel Aviv University, Tel Aviv 69978, Israel

*Equal contribution
[#]Correspondence: roded@post.tau.ac.il and przytyck@ncbi.nlm.nih.gov

## ABSTRACT

The analysis of the mutational landscape of cancer, including mutual exclusivity and co-occurrence of mutations, has been instrumental in studying the disease. We hypothesized that exploring the interplay between co-occurrence, mutual exclusivity, and functional interactions between genes will further improve our understanding of the disease and help to uncover new relations between cancer driving genes and pathways. To this end, we designed a general framework, *BeWith*, for identifying modules with different combinations of mutation and interaction patterns. We focused on three different settings of the *BeWith* schema: (i) *BeME-WithFun* in which the relations between modules are enriched with mutual exclusivity while genes within each module are functionally related; (ii) *BeME-WithCo* which combines mutual exclusivity between modules with co-occurrence within modules; and (iii) *BeCo-WithMEFun* which ensures co-occurrence between modules while the within module relations combine mutual exclusivity and functional interactions. We formulated the *BeWith* framework using Integer Linear Programming (ILP), enabling us to find optimally scoring sets of modules. Our results demonstrate the utility of *BeWith* in providing novel information about mutational patterns, driver genes, and pathways. In particular, *BeME-WithFun* helped identify functionally coherent modules that might be relevant for cancer progression. In addition to finding previously well-known drivers, the identified modules pointed to the importance of MTOR and MED23 genes as well as the interaction between NCOR and NCOA3 in breast cancer. Additionally, an application of the *BeME-WithCo* setting revealed that gene groups differ with respect to their vulnerability to different mutagenic processes, and helped us to uncover pairs of genes with potentially synergetic effects, including a potential synergy between mutations in TP53 and metastasis related DCC gene. Overall, BeWith not only helped us uncover relations between potential driver genes and pathways, but also provided additional insights on patterns of the mutational landscape, going beyond cancer driving mutations.

INTRODUCTION

The analysis of the mutational landscape of cancer has been instrumental in studying the disease and identifying its main drivers and subtypes. In particular, mutual exclusivity of mutations in cancer drivers has recently attracted a lot of attention. This relation can help identify cancer drivers, cancer-driving pathways, and cancer subtypes [1–7]. Although less studied, co-occurrence of mutations has also provided critical information about possible synergistic effects between pairs of genes [8–10].

Importantly, both properties can arise due to several different reasons, making the interpretation of the implied gene-gene relations challenging. Specifically, mutually exclusive mutations *within* functionally interacting genes may indicate that a mutation in either of the two genes dysregulates the same pathway. On the other hand, mutually exclusive mutations might also reflect a situation where two genes drive two different cancer types or subtypes. Such type-specific mutations are more likely to occur *between* genes belonging to different pathways. We have previously observed that within cancer type mutual exclusivity is more enriched with physically interacting pairs of genes compared to the between cancer types mutual exclusivity [3]. Thus, the presence or absence of interactions between mutually exclusive genes might provide hints toward the nature of the mutual exclusivity. In addition, the property of mutual exclusivity of mutations is not necessarily limited to cancer drivers, and therefore a proper understanding of this property is critical for obtaining a better picture of cancer mutational landscape in general and for utilization of mutual exclusivity in cancer driver prediction.

As with mutual exclusivity, co-occurrence of mutations might emerge due to a number of different causes. Perhaps the most important case is when the disabling of two genes simultaneously might be beneficial for cancer progression. Examples of such a scenario include the co-occurrence of TP53 mutation and Myc amplification [9,11] or co-occurring mutations in PIK3CA and RAS/KRAS [5,8,12]. Alternatively, co-occurrence of somatic mutations might indicate the presence of a common mutagenic process. For example, we observed in the previous work that the co-occurrence of TP53 and TTN mutations in breast cancer patients was not statistically significant based on a test corrected with the samples' mutation frequencies, although they were found to co-occur using an uncorrected Fisher's exact test [13]. This suggests that the co-occurrence of TP53 and TTN was more likely due to a common mutagenic process acting on them rather than a benefit to cancer progression.

Given the diversity of reasons for observing mutual exclusivity and co-occurrence relation, we hypothesised that jointly considering co-occurrence, mutual exclusivity and functional interaction relationships will yield a better understanding of the mutational landscape of cancer. As a step in this direction our goal was to develop an approach able to identify groups of genes (or gene modules) that show a coherent with genes inside and outside modules. While many methods to identify cancer related modules exist, such modules are typically identified by focusing on relationships of genes within a module. In particular, there have been several previous attempts to combine mutual exclusivity and functional interactions for module identification [1–3,14]. However, these methods were primarily focused on finding functional modules that include mutually exclusive genes rather than uncovering the relations between such modules.

To address this challenge, we designed a general framework, named *BeWith*, for identifying modules with different combinations of mutation and interaction patterns. On a high level, *BeWith* tackles the following problem: given a set of genes and two types of edge scoring functions (within and between scores), find clusters of genes so that genes within a cluster maximize the "within" scores while gene pairs spanning two different clusters maximize the "between" scores. We formulated the *BeWith* module identification problem as an Integer Linear Programming (ILP) problem and solved it to optimality. The flexibility of the ILP formulation allowed us to include additional constraints such as module connectivity to enhance the module discovery process. To our best knowledge, our work is the first to design a *general framework* that combines mutual exclusivity, co-occurrence, and functional interactions and that is systematically applied to identify sets of modules having both between and within properties.

**Figure 1.** Overview of three settings for which BeWith was applied. (A) The goal of BeME-WithFun is to discover modules which have dense functional interactions within the modules while having mutually exclusive mutations with genes outside the modules (B) In BeME-WithCo we aim to identify modules which have co-occurring mutations within the modules while having mutual exclusivity between the modules (C) In BeCo-WithMEFun, we look for modules of functional and mutual exclusivity relations inside with co-occurring mutations between modules.

In this work, we focused on three different settings of the *BeWith* framework (Figure 1). <u>Setting 1 *(BeME-WithFun)*</u>: Our first setting aims to identify multiple dysregulated pathways using functional interaction and mutual exclusivity information (Figure 1A). Different from the previous module detection methods, we start with the general assumption that genes that are mutually exclusive with some other genes are more likely to be relevant to cancer progression whether or not their partners are in the same pathway. Based on this assumption, we identify modules that are functionally related within and at the same time show mutual exclusivity between the modules. While genes within a module may also be mutually exclusive with each other, we do not optimize for within module mutual exclusivity.

<u>Setting 2 *(BeME-WithCo)*</u>: This setting searches for groups of genes that are co-mutated (within) and, at the same time, are mutually exclusive with genes in other groups (Figure 1B). While in this setting we are mostly interested in co-occurrence within modules, mutual exclusivity between modules helps us pick the gene groups that are potentially relevant to cancer (as in the first setting). There are two possible explanations for co-occurrence of mutations. First, such co-mutations might benefit cancer progression. For example, deficiency in the DNA damage repair machinery by itself does not cause cancer but rather makes a cell vulnerable to cancer causing mutations. Interestingly, if a co-mutation is indeed beneficial for cancer progression, then the mutations in turn might lead to a pattern of mutual exclusivity with other cancer driving mutations as we look for in this setting. Second, co-mutation can arise because a mutational process might affect a subset of genes disproportionately. Indeed, cancer patients show mutational signatures due to a variety of mutagenic processes such as aging, smoking, deficiency of DNA damage process, and APOBEC activity [15,16]. In this context, ensuring mutual exclusivity between modules can help separate groups of genes affected by different mutagenic processes. In either case, combining co-occurrence within modules and mutual exclusivity between modules can lead to a more insightful understanding of the cancer mutational landscape. The setting is another novel way of analyzing the patterns of cancer mutations, using co-occurrence within modules.

*Setting 3 (BeCo-WithMEFun):* Our third setting is specifically designed to identify co-occurring driver pathways. We seek modules displaying functional and mutual exclusivity relations inside a module and co-occurrence between modules (Figure 1C). While Leiserson et al. presented an anecdotal example of such co-occurring pair of modules (of two genes in each) in their GBM data analysis [5], such modules are hard to find.

We applied *BeWith* in these three complementary settings to two TCGA datasets: somatic mutation profiles in breast cancer (BRCA) and endometrial cancer (UCEC). In the main text we focus on the results with the BRCA dataset and defer the results with the UCEC data to Supplementary Material S6.

## RESULTS

### Method Evaluation

To validate our method, we computed the significance of the results compared to those obtained with 100 randomized instances (Supplementary Materials S2). We evaluated the modules obtained in each setting for the objective function value and how well the modules identified known driver genes. We stress that BeWith is not specifically targeted toward detecting cancer driving genes but rather searching for gene modules that may expose various cancer related properties including differences in mutational processes. However the modules are still expected to be enriched with cancer related genes. Indeed, we found that our modules significantly outperformed the random ones with respect to all measures (Table 1).

|  | **Features** | **# Known Drivers** | **Driver Enrichment (Hypergeometric test)** | **Objective Function Value** |
|---|---|---|---|---|
| **BeME_WithFun** | Real | 14 | 6.9e-8 | 57.60 |
|  | Random (average) | 3.85 | 0.074 | 9.92 |
|  | Significance | <0.01 | <0.01 | <0.01 |
| **BeMe_WithCo** | Real | 7 | 2.41e-3 | 43.79 |
|  | Random (average) | 2.55 | 0.17 | 10.75 |
|  | Significance | <0.01 | 0.02 | <0.01 |
| **BeCo_WithMeFun** | Real | 2 | 0.056 | 3.07 |
|  | Random (average) | 0.28 | 0.60 | 1.12 |
|  | Significance | 0.04 | <0.01 | <0.01 |

**Table 1**. Comparison of the results of three settings of the BeWith schema on real and randomized data.

Notably, the statistical significance of modules obtained with BeCo_WithMeFun was lower than for the modules obtained with other methods. For the list of cancer drivers, we used a combined list from COSMIC Cancer Gene Census [18] and 138 cancer driver genes from [19].

**Comparison with methods for cancer module discovery**

Our approach is different from most of previous methods for mutated module identification as we focus on finding modules with relations both within and between modules. Given differences in the objectives, we performed the comparison for the purpose of establishing whether modules identified with *BeME-WithFun* have similar enrichment in cancer genes relative to the modules uncovered by other methods despite the fact that modules uncovered with *BeME-WithFun* are optimised to respect different set of relations.

The most comparable approaches are Multi-Dendrix [20], MEMCover [21] and CoMDP [10]. These algorithms seek to find multiple functional modules based on mutational patterns, enforcing mutual exclusivity relation within modules. Multi-Dendrix and MEMCover identify such multiple modules assuming mutations may potentially co-occur between modules but without enforcing it [20,21]. CoMDP [10] attempted to ensure co-occurrences between the modules. Table 2 shows the comparison of our results in *BeME-WithFun* setting with the modules obtained from Multi-Dendrix and MEMCover using BRCA somatic mutation dataset. Multi-Dendrix looks for multiple modules by optimizing mutation coverage and mutual exclusivity, by which they implicitly aims to ensure functional similarity within the modules. MEMCover optimizes mutation coverage while utilizing functional interactions and mutual exclusivity within modules. Unlike the previous methods, *BeME-WithFun* insists on mutual exclusivity *between* modules while using functional interactions within modules. We set Multi-Dendrix to produce the same number of modules as *BeME-WithFun* and the core modules (combining the results with the maximum module size varied from 2 to 5) were used for comparison. For MEMCover, we obtained modules by setting each patient to be covered at least by 5 mutated genes and then chose the same number of modules as *BeME-WithFun* based on the best coverage. Table 2 shows that *BeME-WithFun* finds better or comparable modules in terms of cancer driver enrichment. As we explicitly enforce functional interactions within modules, the *BeME-WithFun* modules are more functionally coherent as expected.

| Features | #Known Drivers | Driver Enrichment (Hypergeometric test) | Functional Coherence (Distance) |
|---|---|---|---|
| **BeME_WithFun** | 14 | 6.9e-8 | 1.03 |
| **Multi-Dendrix** | 9 | 1.1e-3 | 2.52 |
| **MemCover** | 13 | 3.63e-7 | 1.08 |

**Table 2**. Comparison of module properties obtained with *BeME-WithFun,* Multi-Dendrix and MEMCover on breast cancer data.

CoMDP considers the setting similar to BeCo-WithMEFun but their original results with CNV data identified co-occurring genes that can be attributed to insertion/deletion events in the same locus rather than to co-occurring pathways. With BRCA somatic mutation data and requiring the same number of genes as returned by our algorithm, CoMDP produced two modules: (TP53) and (TTN, USH2A). These included only one known cancer driver (TP53, p=0.33) compared to two drivers (p=0.056) obtained by BeCo-WithMEFun (Section 3.2.3). Importantly, neither TTN nor USH2A significantly co-occur with TP53 after correcting for patient mutation frequencies, making these modules hard to interpret.

In summary, although BeWith is designed to identify gene modules with specific mutation patterns in cancer rather than to find cancer driving genes, the comparison with module finding approaches revealed that BeWith performed well in finding driver genes too.

**BeME-WithFun: Functional modules with mutual exclusivity between modules**

In this setting, we search for functionally related groups of genes with potential relevance for cancer. In order to ensure that genes within each module are likely to be in the same pathway, we enforce functional edges within modules while penalizing functional interactions of genes between different modules. We also reward the edges between modules for mutual exclusivity. By applying this setting to TCGA BRCA somatic mutation dataset, we identified seven modules (Figure 2A) including many prominent drivers in breast cancer such as TP53, AKT1, CDH1, PIK3CA, GATA3, and MAP3K1 (in modules 1, 2, 3 and 6). Notably, most of the mutual exclusivity relations between modules we identified are among different pathways[1] (Figure 2B). The results reveal that mutual exclusivity, although commonly sought for within pathways, may also frequently occur between pathways. In addition, since BeME-WithFun strictly enforces functional interactions within the modules, we obtain functionally coherent modules. In particular, TP53 and GATA3 are typically put together into one mutual exclusivity module [4,5] even though they are only distantly functionally related. Overall, by the method design, the BeME-WithFun modules are densely connected (distance=1.03) by functional interactions as compared with an average distance of 2.52 for Multi-Dendrix and 1.08 for MEMCover on BRCA mutation dataset. The method also allows us to identify less frequently mutated drivers (having weak mutual exclusivity signals) such as MTOR, MED23, FOXA1, PIK3R1 benefiting from the combined analysis with functional interactions.

**Figure 2**. (A) Modules uncovered by BeME-WithFun for TCGA breast cancer dataset. Cyan and brown edges represent pairs of genes with significant mutually exclusive and co-occurring mutations, respectively. Darker edges correspond to lower p-values and p-values less than or equal to 1e-10 have same darkness. Black edges represent functional interactions. The node sizes of genes reflect the number of mutated samples. (B) WeSME p-values between the identified modules. There are many significant mutual exclusivity edges between modules and for some, the significance of mutual exclusivity is increased compared to the ones with gene pairs.

Interestingly, module 7 contains two co-occurring genes: NCOR2 (nuclear co-repressor) and NCOA3 (coactivator), with the latter being a well-known cancer driver [19]. TBLR1 is another nuclear co-repressor and MED23 is a component of the mediator complex and a coactivator involved in regulated transcription. The module was not detected by previous methods probably due to the co-occurrence between NCOR2 and NCOA3 as most of previous methods enforce mutual exclusivity within modules.

---

[1] Except module 1 and 2, in which the genes are closely related and belong to PIK3CA/AKT1/MTOR pathway. They are split into two groups as we maximize mutual exclusivity between modules and the strong mutual exclusivity between modules outweighed the functional relationships.

Although not including known cancer drivers, module 5 contains a cluster of 4 mucins - members of a family of large proteins which are components in most gel-like secretions and some are involved in signalling. Mucins, most predominantly MUC4, have been associated with cancer, typically via abnormal expression [22–24]. However, the source of mutual exclusivity and the role of mutations in these genes are still not clear. Both module 4 and 5 contain many long genes and were also identified in the UCEC dataset (see Supplementary Material S6).

We also examined whether the modules identified by BeME-WithFun are more significantly mutually exclusive with other genes as compared to the mutual exclusivity of the module's individual genes. The confidence of mutual exclusivity test is largely limited by the number of mutated samples, causing the patterns in rarely mutated genes hard to observe. Merging genes in the same module into one supergene[2], we computed WeSME $p$-values between the supergenes and other individual genes, which allowed us to identify many new mutually exclusive pairs. For example, module 2 that contains PIK3CA and CDH1 is mutually exclusive with several genes implicated in cancer including , MED23, and DCC. The mutual exclusivity of these two genes with either PIK3CA or CDH1 was not statistically significant, but is statistically significant with the supergene corresponding to module 2. Interestingly, module 5 (with Mucins) is mutually exclusive with many cancer drivers including PIK3CA, MAP3K1 and RUNX1. The list of all statistically significant module-gene pairs where the statistical significance of the mutual exclusivity of module-gene pair is higher than mutual exclusivity of the given gene with any gene in the module is provided in Supplementary Material S7.

**BeME-WithCo: co-occurrence modules that are mutually exclusive with each other**

In order to identify informative co-occurrence modules, we applied BeME-WithCo and obtained six modules of genes (Figure 3A). As expected, the analysis of the modules in this setting revealed both types of co-occurring modules: modules containing putative cancer drivers with synergistic mutations and modules that are likely a result of common mutagenic processes. Interestingly, we found some modules with both properties, meaning that the genes in the module undergo similar mutational processes but their synergistic roles in cancer were also implicated in the literature.

**Figure 3.** (A) Modules uncovered by BeME-WithCo on breast cancer data. Edge color-coding and node size coding are the same as in Figure 2. (B) Decomposition of the observed mutational spectra of modules 1, 3, and 6 into predefined COSMIC signatures of mutational processes identified in breast cancer [25]. Signature 2 is APOBEC related, Signature 3 is associated with failure of DNA double-strand break-repair and also with BRCA1 and BRCA2 mutations. Signatures 6 and 26 are associated with defective mismatch repair. The aetiology of signatures 5 and 30 is unknown.

Specifically, we found consistent mutational signatures in modules 1, 3, and 6[3] (Figure 3B; see Supplementary Material S4). Mutational signatures are distinctive patterns in mutational spectrum that can reveal the underlying mutation generating processes [15,16]. It is interesting to see that different modules have different composition of mutational signatures, which in turn implies that genes in different modules are affected by different mutagenic processes. The mutual exclusivity between modules in this setting facilitates the partitioning, if present in data. Uncovering such relation is important for a proper interpretation of mutual exclusivity which can be extended to genes beyond cancer drivers. Note that in this setting we used the hypergeometric test for co-occurrence to allow, in particular, detection of modules due to undergoing the same mutagenic processes. We also applied the more stringent WeSCO test for the identified modules to test whether co-occurrence within modules is likely to be functional (see the detailed discussion for three representative modules presented below).

---

[2] For the newly created supergene, we define that the supergene has a mutation in a patient if there is any mutated gene in the module for the patient. WeSME p-values were computed for the supergene and all other genes.
[3] The remaining modules either did not have sufficient number of observed somatic mutations and/or their mutational spectrum could not be decomposed into signatures with a small error (e.g. due to selection towards specific mutations).

For module 1, the mutational signatures associated with DNA repair are the dominating signatures in this module (Signatures 3, 6, and 26). In addition, both genes in the module (TP53 and DCC) are known to be associated with cancer. TP53 is involved in DNA repair, growth arrest, and apoptosis. In particular, mutations in TP53 can lead to uncontrolled proliferation and invasive growth. On the other hand, DCC is suggested to have an anti-metastatic role [26], meaning that DCC may only contribute to cancer in the context of a preexisting condition. We conjecture that the mutations in DCC may be contributing to cancer progression for the patients with defective mismatch repair and/or impaired TP53 functionality.

In contrast, module 3 is most strongly enriched with Signature 3, which is known to be associated with BRCA1 and BRCA2 germline and somatic mutations. The presence of BRCA2 in this module is consistent with the finding. Interestingly, the module includes PREX2, which has been recently identified as a negative regulator of PTEN in breast cancer [27]. In addition, the gene has been shown to be not only significantly mutated in human melanomas but also relevant for melanoma tumorigenesis by a combination of mutations and overexpression [28]. However, the precise mechanism(s) of action remains unknown. The inclusion of PREX2 in cluster with BRCA1/2 mutation pattern might shed some light on possible synergistic interactions of this recently proposed driver.

Different from the above two modules, module 6 contains three long genes including MUC16 and TTN. An interesting aspect of this cluster is the presence of APOBEC related signature (Signature 2) but no mismatch repair associated signatures. For more discussion of this and the remaining modules see Supplementary Material S5.

**BeCo-WithMEFun: mutually exclusive modules that are co-occurring with each other**

Complementing the above analyses, we utilized BeCo-WithMEFun to look for modules that contain mutually exclusive and functionally related genes with co-occurrence between modules. The setting is motivated by the fact that a single mutation may be enough to cause pathway dysregulation (thus mutual exclusivity within a module) and multiple dysregulated pathways are required for cancer progression. We identified a pair of such modules: Module 1 with TP53 and BRCA2, and Module 2 with DCC (Supplementary Fig S4).

Both BRCA1 and BRCA2 are known to interact with TP53, and contribute to DNA repair and transcriptional regulation in response to DNA damage [29,30]. The activation of TP53 can also occur in response to DNA damage amongst other stresses. As discussed in the previous section, DCC is believed to have anti-metastatic role and its reduced functionality is might have a synergistic effect with other cancer driving events. This observation is consistent with the finding in the BeME-WithCo setting, and points to a possible synergy between DCC and broader DNA repair pathway. However, BeCo-WithMEFun did not find larger co-occurring modules in neither of the two cancers types.

**DISCUSSION**

We introduced the BeWith framework to identify multiple mutated modules displaying specific mutation patterns between and within modules. In this work, we considered three settings: BeME-WithFun (ensuring mutual exclusivity of mutations between different modules and functional similarity of genes within modules), BeME-WithCO (ensuring mutual exclusivity between modules and co-occurrence of mutations in genes within modules), and BeCo-WithMEFun (ensuring co-occurrence between modules while enforcing mutual exclusivity and functional interactions within modules). By utilizing these different settings of within and between properties, BeWith revealed complex relations between mutual exclusivity, functional interactions, and co-occurrence. In particular, BeME-WithFun identified functionally coherent modules containing cancer associated genes, including previously unappreciated modules such as the NCOA3/NCOR2 module. Different from most of previous methods focusing on mutual exclusivity within modules, our methods enforce mutual exclusivity between modules. Interestingly, our modules include many known cancer drivers (more or comparable compared to previous methods) while they also exhibit significant mutual exclusivity relationships between modules. The BeME-WithCo setting also allowed us to investigate mutated modules in a novel way by looking for co-occurring mutations inside a module. This setting was particularly insightful in helping us uncover pairs of genes with likely synergetic effects in breast cancer. Going beyond cancer driving mutations, the setting provided additional insights into underlying mutagenic processes in cancer. Specifically, it revealed that different gene groups might differ with respect to their vulnerability to different mutagenic processes. The differences can contribute to strong mutual exclusivity signals between modules. Finally, while using BeCo-WithMEFun, we have been able to elevate some of the observations obtained by BeME-WithCo to the pathway level, the setting did not uncover any larger co-occurring functional modules where the members of individual modules are mutually exclusive. The observation suggests that after conservative correction in co-mutation test with mutation rates, co-occurrence of mutations in two different functional modules appears to be a rather rare event.

Overall, BeWith can be used to uncover relationship between genes, gene groups, and pathways that were not accessible to previous methods. Importantly, the BeWith formulation is very general and can be used to interrogate other aspects of the mutational landscape by exploring different combinations of within-between definitions and constraints with simple modifications.

**METHODS**

We start by defining the Between-Within module finding (BeWith) problem then formulate it as an integer linear program. The optimization problem provides a general framework for identifying a set of clusters. By adjusting reward and penalty functions, and some of the constraints we can apply the framework to detect modules occurring in the different settings described above. In detail, we are given two weight functions *between*(i, j) and *within*(i, j) for pairs of genes between and within modules, respectively. We aim to identify a set of modules which maximizes *between weights* for gene pairs from different modules while maximizing *within weights* inside a module simultaneously. The optimization problem is NP-hard (as it generalizes Max Cut), but can be solved optimally for current datasets as we demonstrate below.

**General ILP Formulation for the Between-Within Module Finding Problem**

Let $K$ be the target number of modules, let $M$ be the maximum number of genes per module, and let $V$ be the set of genes we consider. We aim to group genes into one of $K$ modules, where the $(K+1)$-th cluster includes all unselected genes. Denote $K' = K+1$. We use the binary variable $y_{ik}$ to indicate whether gene $i$ is in module $k$ ($y_{ik} = 1$) or $y_{ik} = 0$ otherwise. We define $ij$ to be a between module pair if gene $i$ and $j$ are in two different modules $k_1, k_2$ respectively ($1 \leq k_1 \neq k_2 \leq K$) and to be a within module pair if both genes belong to the same module $k$ ($1 \leq k \leq K$); $ij$ is an unselected pair otherwise. Additionally, the following integer binary variables are used to capture different types of pairs:

- $x_{ijk} = 1$ if genes $i, j$ are in the same module, $0$ otherwise.
- $z_{ij} = 1$ if pair $ij$ is a between module pair, $0$ otherwise.
- $u_{ij} = 1$ if pair $ij$ is unselected, $0$ otherwise.

The objective of ILP is defined as:

$$Max \ \sum_{ij} between(i,j) \, z_{ij} + \sum_{ij} \sum_{k=1}^{K} within(i,j) \, x_{ijk} \tag{1}$$

The constraints (2)-(3) ensure that each gene $i$ belongs to exactly one of the modules and that the module size is bounded by $M$.

$$\sum_{k=1}^{K+1} y_{ik} = 1 \qquad\qquad \forall i \in V \tag{2}$$

$$\sum_{i \in V} y_{ik} \leq M \qquad\qquad \forall k \in [1, K] \tag{3}$$

The set of constraints (4)-(6) ensure that $x_{ijk} = 1$ if both $i, j$ are selected to module $k$, $1 \leq k \leq K$.

$$x_{ijk} \leq y_{ik} \qquad\qquad \forall ij, \ \forall k \in [1, K] \tag{4}$$
$$x_{ijk} \leq y_{jk} \qquad\qquad \forall ij, \ \forall k \in [1, K] \tag{5}$$
$$x_{ijk} \geq y_{ik} + y_{jk} - 1 \qquad\qquad \forall ij, \ \forall k \in [1, K] \tag{6}$$

Similarly, the constraints (7)-(10) ensure the proper assignment of $u_{ij}$ and $z_{ij}$.

$$u_{ij} \geq y_{iK'} \qquad\qquad \forall ij \tag{7}$$
$$u_{ij} \geq y_{jK'} \qquad\qquad \forall ij \tag{8}$$
$$u_{ij} \leq y_{iK'} + y_{jK'} \qquad\qquad \forall ij \tag{9}$$

$$z_{ij} = 1 - u_{ij} - \sum_{k=1}^{K} x_{ijk} \qquad \forall ij \qquad (10)$$

In some settings, we also added additional constraints for ensuring that the density of the modules is at least $D$:

$$\sum_{j \in \{j' | \exists ij' \in E_X\}} y_{jk} \geq D(M-1)(y_{ik}-1) + D(\sum_{j \in V} y_{jk} - 1) \qquad \forall i \in V \ \forall k \in [1, K] \qquad (11)$$

where $E_X$ is a subset of gene pairs depending on the setting. Note that if $D \geq 0.5$, the module is a connected subgraph since for any two non-adjacent vertices, they must have a common neighbor. We additionally required in some settings that for each gene $i$ in a module, it has at least some edges in a certain type of subset of edges $E_Y$ (e.g., mutual exclusivity or co-occurrence) with genes in other modules.

$$\sum_{j \in \{j' | ij' \in E_Y\}} z_{ij} \geq y_{ik} \qquad \forall i \in V \ \forall k \in [1, K] \qquad (12)$$

Finally, although all the variables $y_{ik}$, $x_{ijk}$, $u_{ij}$, $z_{ij}$ are required to be binary, it is sufficient to require the variables $y_{ik}$ to be binary and leave the other variables $x_{ijk}$, $u_{ij}$, $z_{ij}$ continuous in $[0, 1]$, which makes sure that all the variables in the optimal solution are binary but reduces the running time. To improve the efficiency of the method, we implemented a symmetry breaking technique. Symmetry in ILPs not only allows for equivalent solutions but can create multiple equivalent subproblems in branch-and-bound trees. These equivalent solutions and equivalent subproblems can lead to significant increase in the running time and memory usage of branch-and-bound algorithms. We reduced the symmetry in solving our ILPs by adding constraints to restrict to feasible solution set. For the details of our solution and its impact on the running time we refer to Supplementary Material S3.

**Application of BeWith to TCGA datasets**

We applied BeWith in three complementary settings for two TCGA datasets: somatic mutation profiles in Breast cancer (BRCA, see results below) and endometrial cancer (UCEC, see Supplementary Material S6).

With somatic mutation profiles of 665 BRCA samples (after removing ultra mutated samples), we first computed their mutual exclusivity and co-occurring relationships for each gene pair and constructed networks retaining only significant relationships. To this end, we used WeSME and WeSCO, which are efficient, weighted sampling based methods for testing mutually exclusivity and co-occurrence respectively, taking into account mutation frequencies of cancer samples [13]. Specifically, we constructed a mutual exclusivity network $G_{ME} = (V, E_{ME})$, in which $E_{ME}$ is a set of gene pairs that have significantly mutually exclusive mutations based on WeSME. A co-occurrence network $G_{CO} = (V, E_{CO})$ was computed with a hypergeometric test or a WeSCO test depending on the setting. In addition, we utilized functional interaction information in some settings, obtained from the STRING database [17].

The weights $w_F(ij)$, $w_{ME}(ij)$ and $w_{CO}(ij)$ for each pair are defined based on the protein functional interaction confidence scores, and *p*-values from mutual exclusivity test and co-occurrence test, respectively. The weights are set to be 0 if the edge does not exist. In the three BeWith settings we used different definitions of between and within functions, and slightly different variants of the constraints. See Supplementary Material S2 for details on the scoring functions and parameters.

**ACKNOWLEDGMENTS**

This work was supported by the Intramural Research Program of the National Institutes of Health, National Library of Medicine. RS was partially supported by a Naomi Kadar award. The study was initiated while TMP and RS were participating in the Algorithmic Challenges in Genomics program of the Simons Institute for the Theory of Computing, and continued during the sabbatical stay of RS at NCBI.

# BeWith: A Between-Within Method to Discover Relationships between Cancer Modules via Integrated Analysis of Mutual Exclusivity, Co-occurrence and Functional Interactions
## (Supplementary Material)

Phuong Dao[1,*], Yoo-Ah Kim[1,*], Damian Wojtowicz[1], Sanna Madan[2]
Roded Sharan[3#], and Teresa M. Przytycka[1#]

[1] National Center of Biotechnology Information, National Library of Medicine, NIH, Bethesda, MD, USA
[2] Department of Computer Science, University of Maryland, College Park, MD, USA
[3] Blavatnik School of Computer Science, Tel Aviv University, Tel Aviv 69978, Israel

*Equal contribution
[#]Correspondence: roded@post.tau.ac.il and przytyck@ncbi.nlm.nih.gov

## S1 Details of different settings for BeWith

### S1.1 BeME-WithFun

This setting searches for functionally related groups of genes with potential relevance for cancer. In order to ensure that genes within each module are likely to be in the same pathway, we enforce functional edges within modules while penalizing functional interactions of genes from different modules. We also reward mutual exclusivity within a module so our optimization function is:

$$Max \sum_{ij \in E_U} \sum_{k=1}^{K} w_F(ij) x_{ijk} + \sum_{ij \in E_U} \left(w_{ME}(ij) - w_F(ij)\right) z_{ij}$$

To strengthen the functional relationships among genes of the same module, we utilize the constraints (11) to ensure that each module is dense in the functional interaction network. In addition, we additionally required that for each gene $i$ in a module, it has at least some mutual exclusivity edge(s) with genes in other modules as in constraints (12) by setting $E_Y = E_{ME}$.

$$\sum_{j \in \{j'|ij' \in E_{ME}\}} z_{ij} \geq y_{ik} \qquad \forall i \in V \; \forall k \in [1, K]$$

### S1.2 BeME-WithCo

In order to identify co-occurring modules, we perform BeWith enforcing the co-occurrence within a module but penalizing within module mutual exclusivity ($within(i, j) = w_{CO}(ij) - w_{ME}(ij)$). To capture co-occurrence modules that are biologically relevant we reward mutual exclusivity relation between modules ($between(i, j) = w_{ME}(ij) - w_{CO}(ij)$). The objective function is then defined as follows:

$$Max \sum_{ij \in E_U} \left(w_{ME}(ij) - w_{CO}(ij)\right) z_{ij} + \sum_{ij \in E_U} \sum_{k=1}^{K} \left(w_{CO}(ij) - w_{ME}(ij)\right) x_{ijk}$$

To strengthen the co-occurrence within each module, we enforce that each module has dense co-occurring interactions by the constraints (11). Similarly to BeME-WithFun, we utilize the constraints (12) to enforce stronger mutual exclusivity requirement ($E_Y = E_{ME}$) among the modules.

**S1.3 BeCo-WithMEFun**

Complementing the above analyses we utilized BeWith to look for modules that contain mutually exclusive and functionally related genes modules that might co-occur with other modules. Specifically, we enforce exclusivity while penalizing co-occurring mutations within modules ($within(i, j) = w_{ME}(ij) - w_{CO}(ij) + w_F(ij)$). Genes in different modules are rewarded for co-occurrence ($between(i, j) = w_{CO}(ij) - w_{ME}(ij) - w_F(ij)$). The objective function is then defined as:

$$Max \sum_{ij \in E_U} \left(w_{CO}(ij) - w_{ME}(ij) - w_F(ij)\right) z_{ij} + \sum_{ij \in E_U} \sum_{k=1}^{K} \left(w_{ME}(ij) - w_{CO}(ij) + w_F(ij)\right) x_{ijk}$$

In order to ensure genes within modules are likely to be in the same pathway, we ensure that each module is a dense subnetwork in STRING functional interaction network using the constraints (11). To strengthen the co-occurrence between the modules and mutually exclusivity within each module, we additionally required that for each gene $i$ in a module, it has at least some co-occurrence edge(s) with genes in other modules ($E_Y = E_{CO}$):

$$\sum_{j \in \{j' | ij' \in E_{CO}\}} z_{ij} \geq y_{ik} \qquad \forall i \in V \; \forall k \in [1, K]$$

## S2. Data and Parameter Selection

Processing mutation data: We utilized TCGA BRCA and UCEC somatic mutation datasets. After removing ultra mutated samples (> 1000 combined alterations), we obtained 665 BRCA samples and 207 UCEC samples. For both datasets, we performed WeSME tests for all gene pairs to estimate the significance of mutual exclusivity. WeSME is an efficient, mutation frequency aware testing [1] for mutual exclusivity. For co-occurrence, we performed both WeSCO and hypergeometric tests depending on settings. Similar to WeSME, WeSCO is a mutation frequency aware test for co-occurrence. It is more stringent than hypergeometric test and therefore can potentially remove some mutational patterns by chance. We used hypergeometric test for BeME-WithCO in which we want to identify mutational signature and performed WeSCO analysis for the identified modules. For both datasets, we consider genes mutated in at least 7 samples.

For the BRCA dataset, we retained 174 pairs of genes with $p \leq 0.05$ from WeSME in all settings. In BeME-WithCo, 1,891 co-occurrent pairs of genes ($p \leq 0.01$, hypergeometric test) were used for further analysis. In BeCo-WithMEFun, we retained 1,235 pairs of genes with WeSCO ($p \leq 0.05$). For the UCEC dataset, we utilized 280 pairs of genes with $p \leq 0.001$ from WeSME and 1,028 CO pairs of genes with $p \leq 0.001$ from hypergeometric tests. To run BeCoWithMEFun on UCEC data, we retain 992 pairs of genes with WeSCO $p \leq 0.05$. To obtain weights for ME/CO edges, for each significant pairs, we compute the weight of the edge as $min(-log_{10}p, 3)$, which sets the weight at least 1 and gradually increased when the p-value gets smaller.

Functional interactions: To obtain the edge weights for functional interactions, we download functional protein interactions from STRING database version 10.0. We use the interactions with high confidence scores (>=900), then divided by the maximum (1000) to obtain the functional interaction edge weight.

The number of clusters: We define $f(k)$ as the value of the objective function by setting the number of modules to $k$. We define $f(k)/k$ as the average module benefit. Starting with $k = 2$ and iteratively increased by 1, we observe that $f(k)/k$ gradually increases but peaks around 4 or 5 then start decreasing. We pick the number of clusters after $f(k)/k$ is maximized. Figure S1 shows that we stops at $k = 7$ for BeME-WithFun, $k = 6$ for BeME-WithCo on BRCA dataset, $k = 6$ for BeME-WithFun and $k = 5$ BeME-WithCo on UCEC dataset. We used $k = 2$ for BeCoWithMEFun with both TCGA BRCA and UCEC datasets.

**Fig. S1.** The number of modules $k$ versus average benefit per module $f(k)/k$ for TCGA BRCA and UCEC datasets.

<u>Density and maximum number of genes per module:</u> We set the density of the modules $D = 0.7$ to make sure that the majority of genes in the same module are functionally interacting. For the maximum number of genes per module , we set $M = 10$ . Note that the maximum number of genes in our modules on both BRCA and UCEC datasets is less than or equal to 5.

<u>Random instances for method evaluation:</u> For each randomized instance, we first randomized the STRING functional network by swapping edges, which preserves node degrees. We also permuted the *p*-values of mutual exclusivity and co-occurrence tests among all pairs with their p-values <=0.25. 100 random instances were generated to compute the significance of BeWith modules.

## S3 Symmetry Breaking for Solving ILP

Symmetry in ILPs can lead to significant increase in the running time and memory usage of branch-and-bound algorithms because it not only allows for equivalent solutions but can create multiple equivalent subproblems in branch-and-bound trees. Previous research have suggested adding constraints to restrict to feasible solution set in order to reduce the symmetry in solving ILPs [2,3].

A feasible solution to our ILP contains the assignments of the variables $y_{ik}$ which assign gene $i$ to module $k$. These assignments correspond to a 0-1 matrix where rows represent genes and columns represent modules. For example, $Y_1$ and $Y_2$ represent two feasible solutions when assigning 4 genes to 3 modules. However, $Y_1$ and $Y_2$ have the same objective value since we can permute the columns of $Y_1$ to obtain $Y_2$.

$$Y_1 = \begin{bmatrix} 1 & 0 & 0 \\ 0 & 0 & 1 \\ 0 & 0 & 1 \\ 0 & 1 & 0 \end{bmatrix}, \quad Y_2 = \begin{bmatrix} 1 & 0 & 0 \\ 0 & 1 & 0 \\ 0 & 1 & 0 \\ 0 & 0 & 1 \end{bmatrix}$$

To reduce the number of equivalent solutions, we restrict our feasible solutions to assignment matrices with columns in increasing lexicographical order. The column/module $M_1$ is lexicographically smaller than column $M_2$ if the smallest index of the genes of $M_1$ is less than that of $M_2$. In the above example, $Y_2$ has columns in increasing lexicographical order while $Y_1$ does not. In $Y_1$, column 3 is lexicographically smaller than column 2.

We can add the following constraints to the ILP to restrict the feasible region to assignment matrices with columns in increasing lexicographical order:

$$\sum_{l=k}^{K} y_{il} \leq \sum_{j=1}^{i-1} y_{j(k-1)} \qquad \forall i \in V \ \forall k \in [2, K], \ i \geq k$$

The above constraints ensure that we only assign gene $i$ to one of the modules $k, k+1, ..., K$ if we already assign one of genes with smaller index to module $k-1$. Similar constraint sets were proposed to allocate surgery blocks to operating rooms [4].

As shown in Fig. S2, the running time of BeWithCluster is significantly improved when adding symmetry breaking constraints for larger $k$. The running time of BeCo-WithMEFun is less than 10 seconds on both BRCA and UCEC datasets. Furthermore, we found no significant difference in results with and without symmetry breaking constraints.

**Fig. S2.** The running time of BeWithCluster with and without symmetry breaking of BeME-WithFun and BeME-WithCo on TCGA BRCA and UCEC datasets.

### S4. Decomposition of Module Mutational Spectrum into Mutational Signatures

We collected all observed single-nucleotide variants in TCGA together with their immediate sequence context for all genes in each module. Mutational spectrum for each module was calculated. We decomposed the mutational spectra into predefined mutational signatures using the R package *deconstructSigs* (version: 1.8.0) [5]. Sanger COSMIC Signatures of Mutational Processes identified in breast cancer [6] were used for decomposition. The input exome data were normalized to the whole genome. Signatures were not extracted for Modules 2, 4, and 5 due to either small number of somatic variants in module genes or large number of presumably selective mutations that lead to increased error in decomposition.

### S5. Application of BeWith to TCGA BRCA datasets

#### S5.1 Additional information about BeME-WithCo modules

This section supplements the discussion of modules in Section 3.2.2 BeME-WithCo: co-occurrence modules that are mutually exclusive each other in the main text (Figure 3).

Module 2 includes NCOR2 (nuclear co-repressor) and coactivator (NCOA3), which were also included in FU Module 7. Their co-occurrence remain statistically significant with more stringent WeSCO test; the mutational spectrum was not decomposed due to small number of somatic mutations in the module. This suggests that the co-occurrence is less likely by chance and the mutations may jointly contribute to cancer progression.

Module 6 contains three long genes including MUC16 and TTN, for which recent studies caution that the frequent mutations might not necessarily be related to cancer progression even though they are often found significantly mutated in cancer [7]. The co-occurrence of TTN and MUC16 is not statistically significant when corrected with WeSCO for patient mutation frequency yet their co-occurrences with GON4L remain statistically significant (Table S1). An interesting aspect of this cluster is the presence of APOBEC related signature (Signature 2) but no mismatch repair associated signatures. This might explain the co-occurrence of TTN and MUC16 and their mutual exclusivity with genes in other modules. While the mutations in TTN and MUC16 are suspected by many to be the results of mutagenic process rather than synergistic mutations in cancer, it is worth noting that MUC16 as well as GON4L can drive cancer growth when overexpressed [8,9]. While overexpression and mutations can have synergistic effects [10] and GON4L is found overexpressed in 26% of TCGA breast cancers, a mechanistic relation of these mutations in cancer progression remains still not clear.

Finally, we point out that the co-occurrence of gene pairs in modules 4 and 5 is not statistically significant when evaluated with WeSCO and thus is most likely related to sample mutation rates.

For BeME-WithCo, we used hypergeometric test to estimate the significance of co-occurrence. After obtaining the modules, we computed, for every pair of genes in each module, several measures of co-mutation including jaccard index and p-value for WeSCO test (Table S1). Unlike hypergeometric test, WeSCO corrects for mutation frequency in each sample.

|  | gene1 | gene2 | jaccard index | #g1 | #g2 | hypergeometric | WeSCO |
|---|---|---|---|---|---|---|---|
| Module 1 | DCC | TP53 | 0.041860465 | 12 | 212 | 0.002505756 | 0.03769 |
| Module 2 | MEF2A | NCOR2 | 0.1875 | 13 | 25 | 2.17E-06 | 4.20E-05 |
| Module 2 | NCOA3 | MEF2A | 0.24137931 | 23 | 13 | 3.34E-08 | 0 |
| Module 2 | NCOA3 | NCOR2 | 0.263157895 | 23 | 25 | 6.52E-10 | 0 |
| Module 3 | BRCA2 | RELN | 0.107142857 | 12 | 19 | 0.003706016 | 0.0158 |
| Module 3 | BRCA2 | PREX2 | 0.142857143 | 12 | 12 | 0.000904382 | 0.00425 |
| Module 3 | RELN | PREX2 | 0.107142857 | 19 | 12 | 0.003706016 | 0.0158 |
| Module 4 | PIK3CA | MAP2K4 | 0.072 | 237 | 31 | 0.0076195 | 0.218 |
| Module 5 | PCDH19 | GATA3 | 0.060240964 | 13 | 75 | 0.009968893 | 0.0667 |
| Module 6 | GON4L | TTN | 0.053571429 | 8 | 110 | 0.000382004 | 0.00485 |
| Module 6 | TTN | MUC16 | 0.12987013 | 110 | 64 | 0.001584299 | 0.135 |
| Module 6 | GON4L | MUC16 | 0.058823529 | 8 | 64 | 0.004077911 | 0.022923 |

**Table S1**. Co-occurrence of genes within modules from BeME-WithCo. #g1 (#g2 respectively) is the number of samples in which gene1 (or gene2) in the pair has been mutated.

### S5.2 Additional information about BeCO-WithMEFun modules

**Fig S3.** Modules uncovered by BeCO-WithMEFun for breast cancer TCGA dataset. Edge color-coding and node size coding are the same as in Figure 2.

### S6. Application of BeWith to TCGA UCEC datasets

### S6.1 Evaluation

We evaluated the results with UCEC dataset compared with random instances and Multi-dendrix using the same method as with BRCA dataset.

|  | Features | # Known Drivers | Driver Enrichment (Hypergeometric test) | Objective Function |
|---|---|---|---|---|
| BeME_WithFun | Real | 10 | 5.1e-4 | 116.08 |

|  | Random (average) | 3.75 | 0.08 | 15.17 |
|---|---|---|---|---|
|  | Significance | <0.01 | 0.03 | <0.01 |
| BeMe_WithCo | Real | 7 | 1.48e-3 | 84.55 |
|  | Random (average) | 1.56 | 0.27 | 13.97 |
|  | Significance | <0.01 | <0.01 | <0.01 |
| BeCo_WithMeFun | Real | 2 | 0.056 | 2.36 |
|  | Random (average) | 0.24 | 0.61 | 1.00 |
|  | Significance | <0.01 | <0.01 | <0.01 |

**Table S2**. Comparison of the results of the three settings of the BeWith schema on real and randomized UCEC data.

| Features | # Known Drivers | Driver Enrichment (Hypergeometric test) | Functional Coherence (Distance) |
|---|---|---|---|
| BeME_WithFun | 10 | 5.1e-4 | 1.0 |
| Multi-Dendrix | 10 | 5.31e-6 | 2.31 |
| MEMCover | 13 | 6.25e-8 | 1.76 |

**Table S3**. Comparison of the results of BeME-WithFun and Multi-Dendrix (UCEC).

### S6.2. BeME-WithFun

By applying this setting to TCGA UCEC somatic mutation dataset, we identified six modules (see figure Figure S4) that include key prominent drivers/associated genes in endometrial cancer such as TP53, CCND1, KRAS, PTEN, PIK3CA, PIK3R1, RPL22, ARID1A. BeME-WithFun also recovered the same TTN and mucin modules as in the case of breast cancer. The emergence of these modules cannot be explained by patient mutation frequency alone as WeSME test take into account the mutation frequencies. Surprisingly, we also retrieved a module related to nonsense mediated decay (SMG1, SMG7). SMG7 is also known to regulate p53 stability and function in DNA damage stress response

**Figure S4**. Modules uncovered by BeME-WithFun for endometrial TCGA dataset. Edge color-coding and node size coding are the same as in Figure 2.

### S6.3 BeME-WithCo

BeME-WithCo retrieved five modules including two one-element modules. Interestingly each of the modules contains at least one driver gene (Figure S5).

**Figure S5**. Modules uncovered by BeME-WithCo for endometrial TCGA dataset. Edge color-coding and node size coding are the same as in Figure 2.

### S6.4. BeCo-WithMEFun

BeCo-WithMEFun identified ATM gene as being co-mutated with an interacting pair containing RPL22 (Ribosomal protein L22) and SMG7 (one of nonsense mediated RNA decay genes). While these three genes have been linked to cancer [11], [12], Kandoth [13], it remains to be establish whether or not there is a synergistic relation between them.

**Figure S6**. Modules uncovered by BeCo-WithMEFun for TCGA UCEC dataset. Edge color-coding and node size coding are the same as in Figure 2.

### S7. WeSME with BeME-WithFun modules

In column *module pv*, we compute the WeSME p-values between the genes and the corresponding modules while the columns *best pairwise p-value* contain the smallest WeSME p-values between the genes and any genes in the modules.

| gene | module pv | best pairwise p-value |
|---|---|---|
| MED23 | 4.40E-05 | 0.000238 |
| F8 | 0.000845 | 0.003535 |
| FREM2 | 0.000845 | 0.003535 |
| TBL1XR1 | 0.00199 | 0.00548 |
| ATM | 0.00215 | 0.01255 |
| DNM3 | 0.00408 | 0.00922 |
| STARD8 | 0.00408 | 0.00922 |
| AOAH | 0.0079 | 0.0084 |
| BRWD3 | 0.00933 | 0.01817 |
| KIF5A | 0.00933 | 0.01817 |
| LETM1 | 0.00933 | 0.01817 |
| GPR112 | 0.00933 | 0.01817 |
| ADD2 | 0.00933 | 0.01817 |
| TRAPPC8 | 0.00933 | 0.01817 |
| GRIK1 | 0.00933 | 0.01817 |
| ZNF438 | 0.00933 | 0.01817 |
| NUP214 | 0.00933 | 0.01817 |
| MED15 | 0.00933 | 0.01817 |
| DNAH1 | 0.01 | 0.0274 |
| BRCA2 | 0.01 | 0.0274 |

**Table S7.1** WeSME mutual exclusivity p-values between FU Module 1 (TP53, MTOR, PTEN, AKT1, PIK3R1) and other genes

| gene | module pv | best pairwise p-value |
|---|---|---|
| USP36 | 7.00E-06 | 2.30E-05 |
| PEG3 | 0.000169 | 0.000425 |
| UTRN | 0.00117 | 0.00278 |
| CROCCP2 | 0.00119 | 0.00525 |

| gene | | |
|---|---|---|
| SPTA1 | 0.00143 | 0.0057 |
| NWD1 | 0.00195 | 0.00366 |
| DDR2 | 0.00195 | 0.00366 |
| HECW2 | 0.00238 | 0.00573 |
| VPS13C | 0.002553 | 0.00573 |
| DYNC2H1 | 0.00263 | 0.00586 |
| RAB3GAP2 | 0.00408 | 0.00739 |
| VCX3B | 0.00408 | 0.00739 |
| RBM23 | 0.00408 | 0.00739 |
| DMBT1 | 0.00408 | 0.00739 |
| MAGEC1 | 0.00408 | 0.00739 |
| ITGA1 | 0.00408 | 0.00739 |
| ZDBF2 | 0.00547 | 0.0114 |
| BCORL1 | 0.00547 | 0.0114 |
| NBPF1 | 0.00676 | 0.00821 |
| COL6A5 | 0.00771 | 0.017147 |
| DNAH8 | 0.0091 | 0.01776 |
| DCC | 0.0091 | 0.01776 |
| ASXL3 | 0.0091 | 0.01776 |
| SMCHD1 | 0.00975 | 0.0139 |
| PRKCQ | 0.00975 | 0.0139 |
| AP1G2 | 0.00975 | 0.0139 |
| CYP2A13 | 0.00975 | 0.0139 |
| LETM1 | 0.00975 | 0.0139 |
| GPR179 | 0.00975 | 0.0139 |
| TNN | 0.00975 | 0.0139 |
| C20orf26 | 0.00975 | 0.0139 |
| PARP8 | 0.00975 | 0.0139 |
| MECOM | 0.00975 | 0.0139 |
| PTCHD2 | 0.00975 | 0.0139 |
| ZNF687 | 0.00975 | 0.0139 |
| TNIK | 0.00975 | 0.0139 |
| EWSR1 | 0.00975 | 0.0139 |
| MED13 | 0.00975 | 0.0139 |
| STAT4 | 0.00975 | 0.0139 |
| AMPD1 | 0.00975 | 0.0139 |

**Table S7.2** WeSME mutual exclusivity p-values between FU Module 2 (PIK3CA, CDH1) and other genes

| gene | module pv | best pairwise p-value |
|---|---|---|
| ATM | 0.0215 | 0.0341 |
| GPR98 | 0.0215 | 0.0341 |
| MDN1 | 0.024 | 0.0381 |
| XIRP2 | 0.024 | 0.0381 |
| PTEN | 0.0211 | 0.0406 |
| MUC4 | 0.0289 | 0.0482 |
| DYNC2H1 | 0.0401 | 0.0551 |
| NEB | 0.0354 | 0.0704 |
| DMD | 0.0354 | 0.0704 |
| COL12A1 | 0.0432 | 0.0756 |
| PCNXL2 | 0.0432 | 0.0756 |
| MED23 | 0.0432 | 0.0756 |
| KIAA1109 | 0.0432 | 0.0756 |
| UBR5 | 0.0432 | 0.0756 |
| MUC17 | 0.0459 | 0.0822 |

**Table S7.3** WeSME mutual exclusivity p-values between FU Module 3 (FOXA, GATA3) and other genes

| gene | module pv | best pairwise p-value |
|---|---|---|
| MAP3K1 | 0 | 0.00003 |
| GATA3 | 1.00E-06 | 0.000067 |
| DCC | 0.00677 | 0.0297 |
| NCOA3 | 0.00718 | 0.009 |
| RYR1 | 0.0122 | 0.015 |
| NR1H2 | 0.0125 | 0.05165 |
| CROCCP2 | 0.0223 | 0.0249 |
| F8 | 0.0239 | 0.0675 |
| MGAM | 0.0285 | 0.0878 |
| FLG | 0.0365 | 0.13 |
| WDR67 | 0.0384 | 0.0867 |
| NFE2L3 | 0.0384 | 0.0867 |
| RP1L1 | 0.0384 | 0.0867 |
| MEFV | 0.0384 | 0.0867 |
| HRNR | 0.0434 | 0.112 |

**Table S7.4** WeSME mutual exclusivity p-values between FU Module 3 (TTN, DMD, NEB) and other genes

| gene | module pv | best pairwise p-value |
|---|---|---|
| MAP3K1 | 0.000181 | 0.0213 |
| PIK3CA | 0.000454 | 0.00451 |
| GATA3 | 0.000597 | 0.00549 |
| PCDH15 | 0.00204 | 0.121 |
| TSC22D1 | 0.0105 | 0.21 |
| RUNX1 | 0.0163 | 0.0319 |
| FAT2 | 0.0192 | 0.121 |
| TBX3 | 0.0258 | 0.179 |
| LRP1 | 0.0317 | 0.162 |
| TEX15 | 0.0317 | 0.162 |
| ODZ2 | 0.0317 | 0.162 |
| MLLT4 | 0.0317 | 0.162 |
| TG | 0.0317 | 0.162 |
| CSMD1 | 0.0371 | 0.122 |
| HERC2 | 0.0391 | 0.168 |
| MLL2 | 0.0391 | 0.0924 |
| HCFC1 | 0.0495 | 0.178 |
| ITPR1 | 0.0495 | 0.178 |
| CDC42BPA | 0.0495 | 0.178 |
| COL14A1 | 0.0495 | 0.178 |

**Table S7.5** WeSME mutual exclusivity p-values between FU Module 5 (MUC4, MUC16, MUC12, MUC5B) and other genes

| gene | module pv | best pairwise p-value |
| --- | --- | --- |
| FAT3 | 0.0381 | 0.15 |
| DOCK11 | 0.0497 | 0.153 |
| RB1 | 0.0497 | 0.153 |
| HERC2 | 0.0497 | 0.153 |
| CCDC66 | 0.0497 | 0.153 |
| SHROOM4 | 0.0497 | 0.153 |
| DNAH10 | 0.0255 | 0.111 |
| DNAH3 | 0.0313 | 0.101 |
| PIK3R1 | 0.0202 | 0.0843 |
| PKHD1L1 | 0.0102 | 0.0576 |
| DMD | 0.0088 | 0.0476 |
| MUC5B | 0.0289 | 0.0662 |
| HRNR | 0.0289 | 0.0662 |
| FLG | 0.0292 | 0.059 |
| RYR2 | 0.0188 | 0.0411 |
| CDH1 | 0.0493 | 0.0675 |
| MUC4 | 0.0278 | 0.0336 |
| MUC16 | 0.0278 | 0.0336 |

**Table S7.6** WeSME mutual exclusivity p-values between FU Module 6 (MAP3K1, MAP2K4) and other genes

| gene | module pv | best pairwise p-value |
| --- | --- | --- |
| TP53 | 1.30E-05 | 0.000282 |
| MUC12 | 0.00105 | 0.0595 |
| USH2A | 0.00419 | 0.1001 |
| RYR2 | 0.0241 | 0.109 |
| NEB | 0.0368 | 0.234 |
| PKHD1L1 | 0.0377 | 0.242 |
| MLL3 | 0.0437 | 0.142 |
| FRG1B | 0.044 | 0.127 |

**Table S7.7** WeSME mutual exclusivity p-values between FU Module 7 (NCOR2, NCOA3, MED23, TBL1XR1) and other genes

Figure 1.

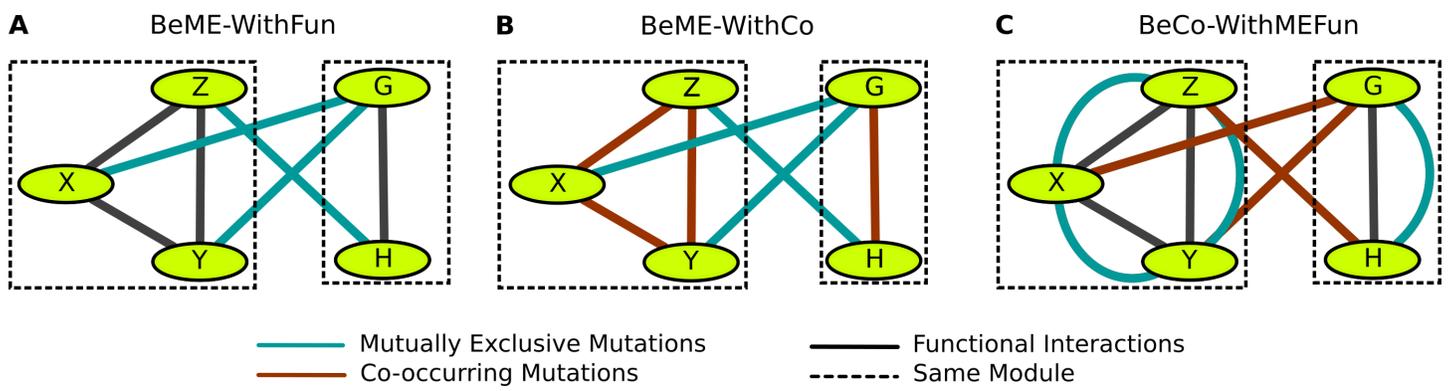

Figure 2.

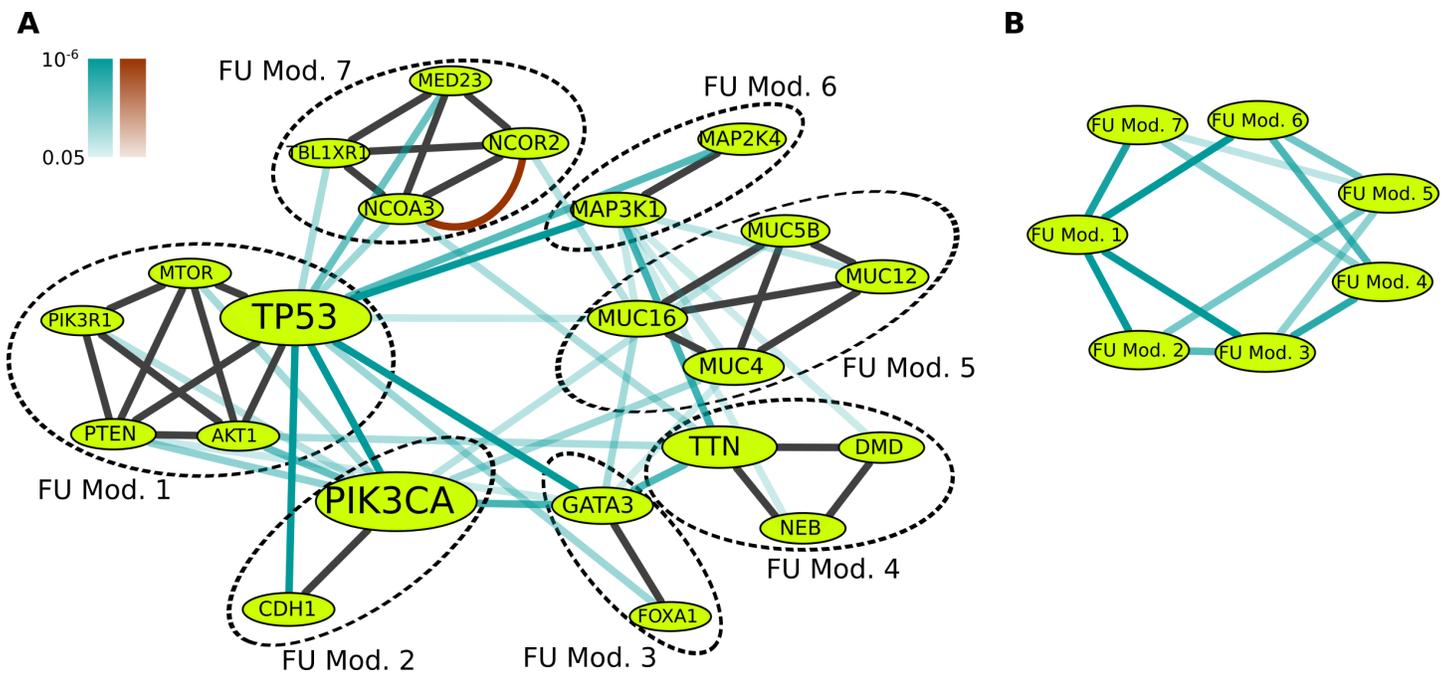

Figure 3.

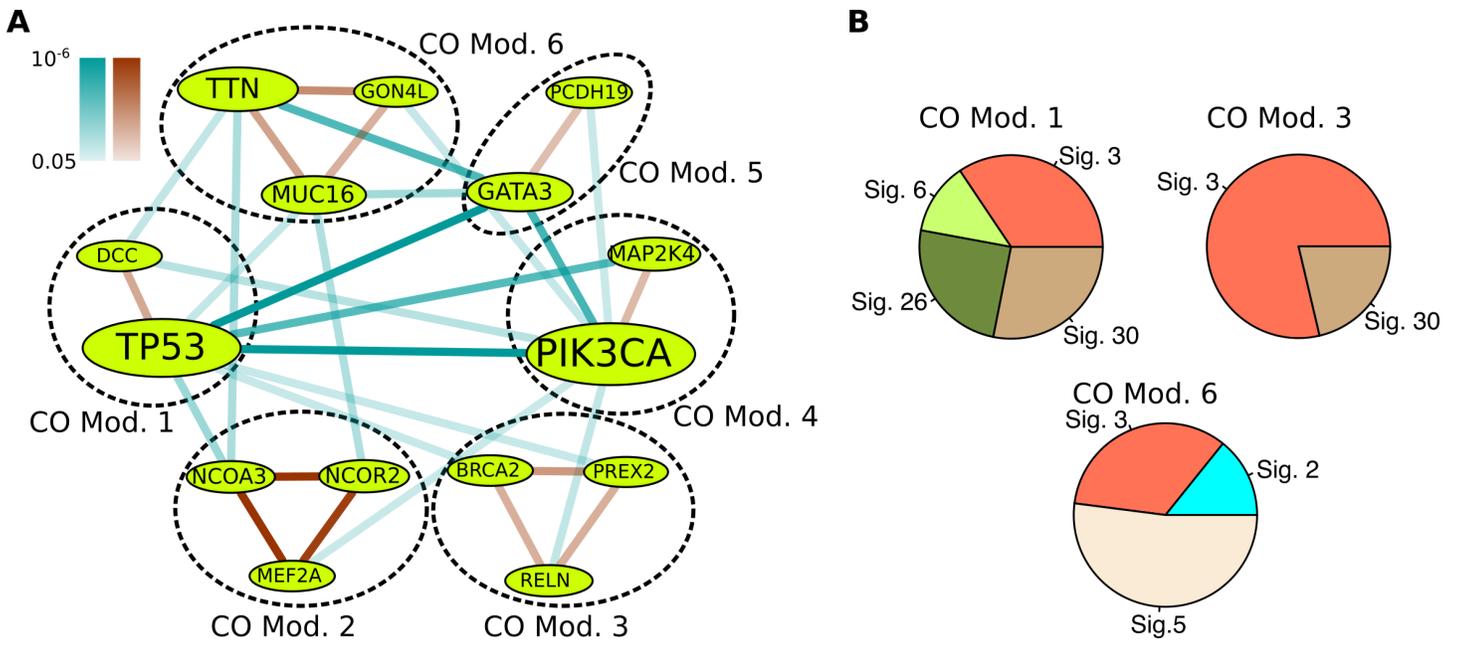

Figure S1.

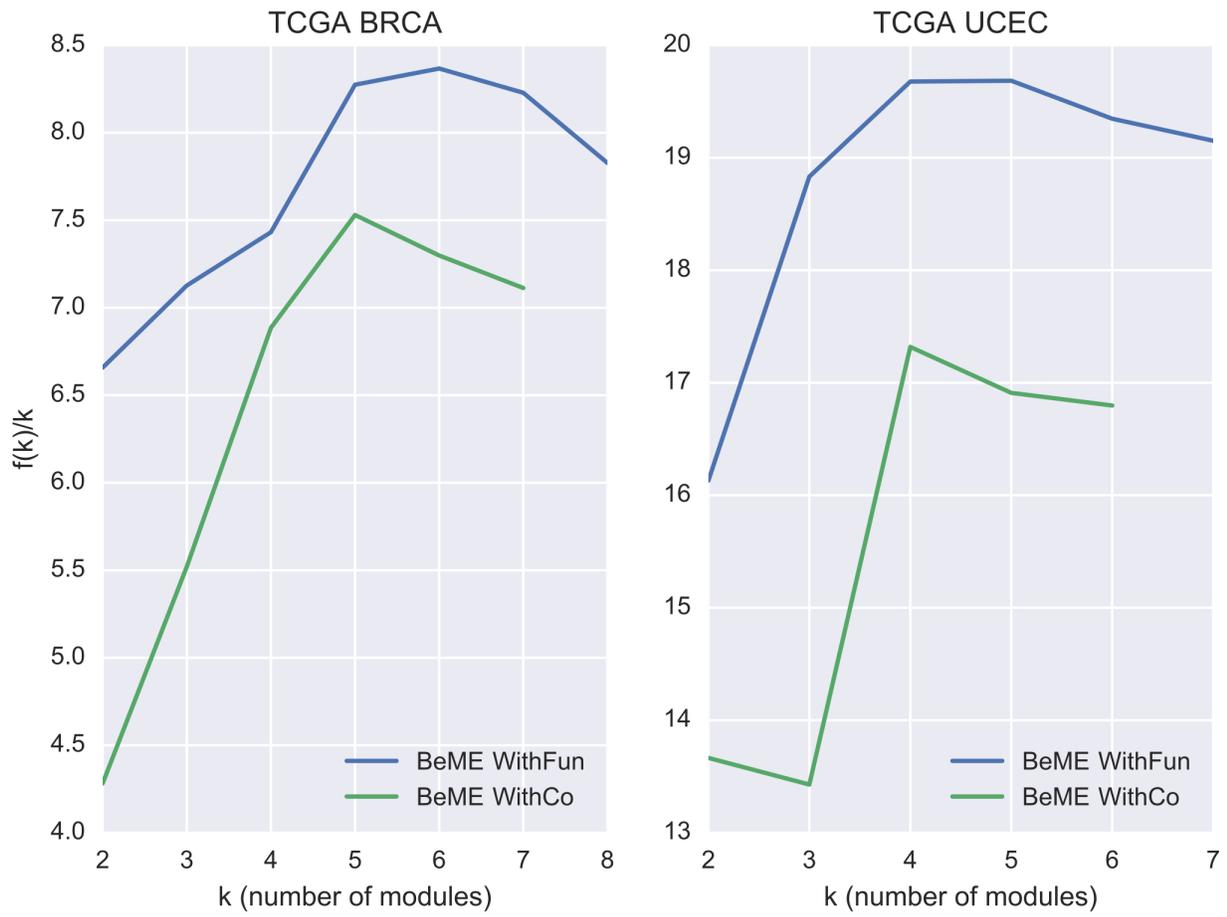

Figure S2.

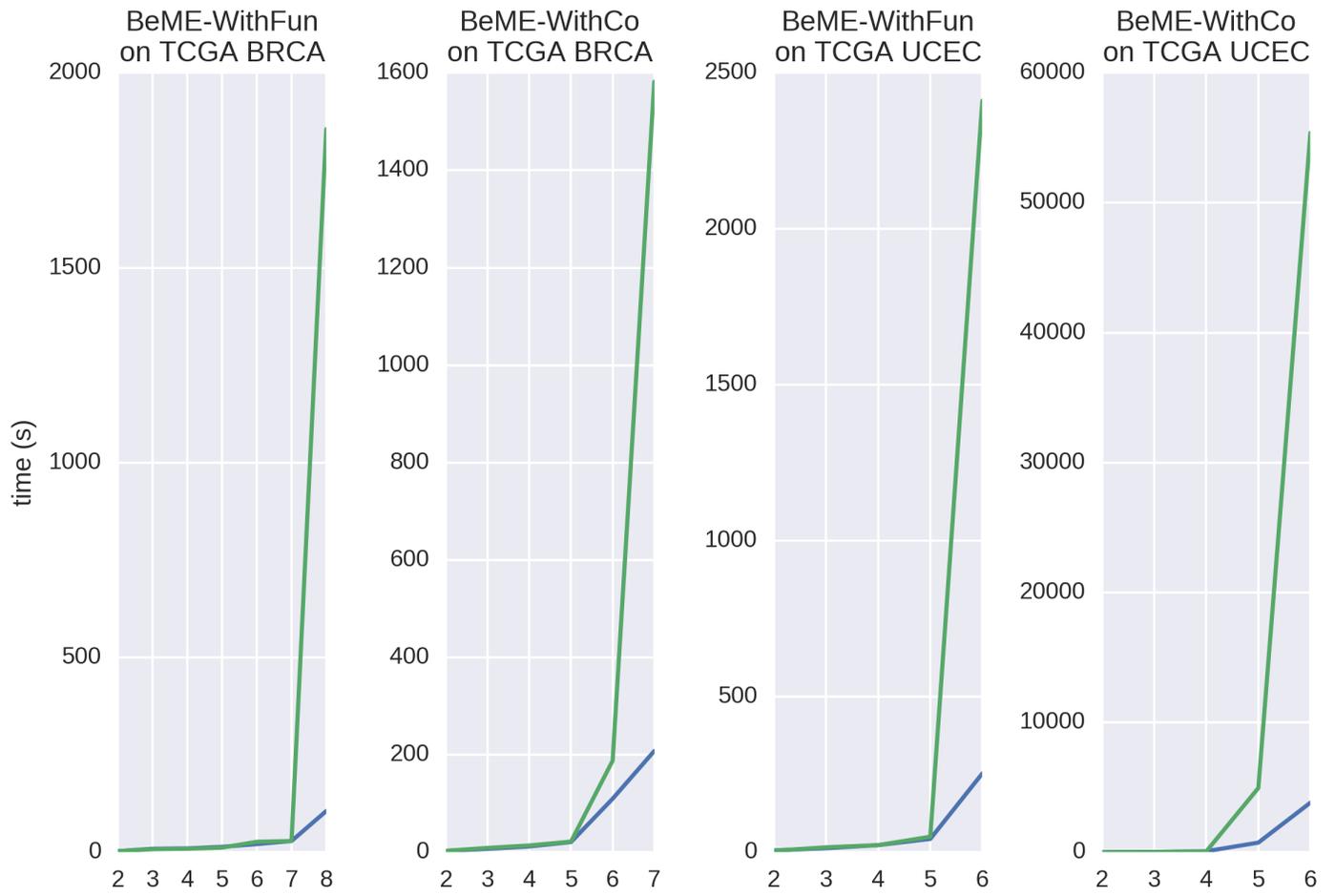

Figure S3.

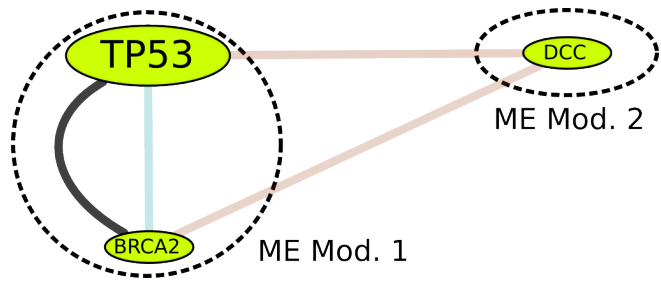

Figure S4.

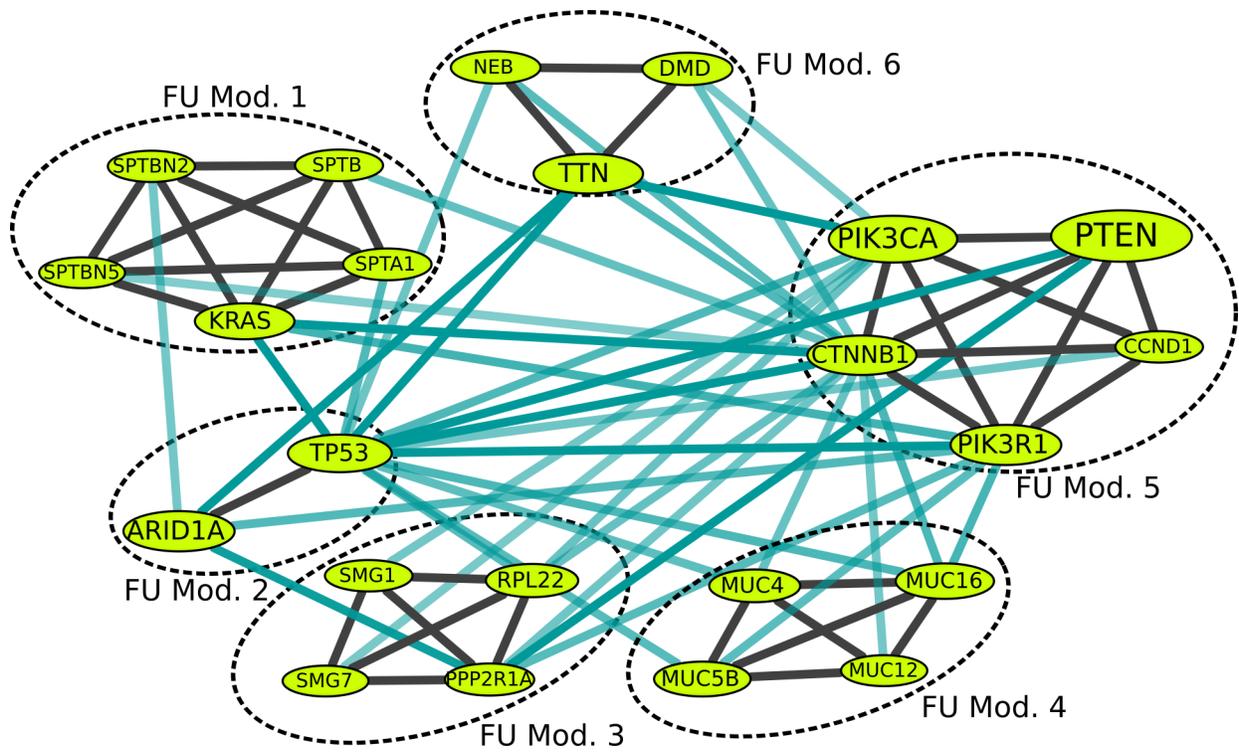

Figure S5.

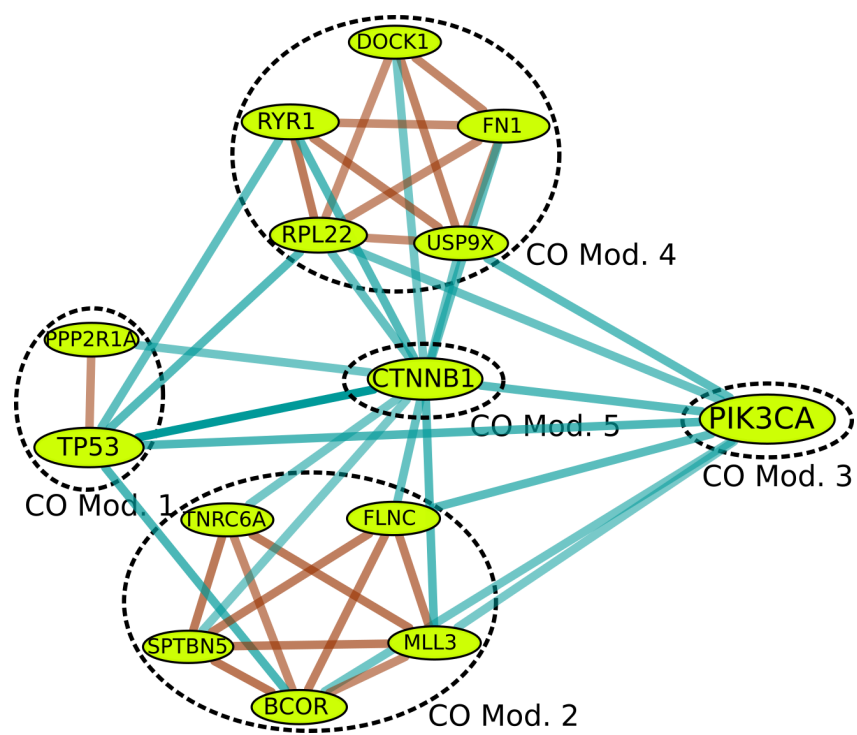

Figure S6.

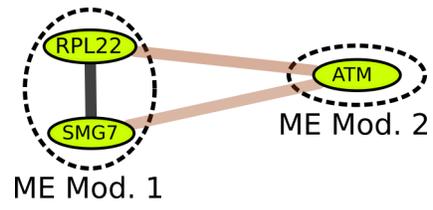